# Room-temperature spin injection and spin-to-charge conversion in a ferromagnetic semiconductor / topological insulator heterostructure

Shobhit Goel,[1,2,#] Nguyen Huynh Duy Khang,[3,4] Le Duc Anh,[1,5,6] Pham Nam Hai,[2,3,7,+] and Masaaki Tanaka[1,2,7,†]

[1]*Department of Electrical Engineering and Information Systems, The University of Tokyo, 7-3-1 Hongo, Bunkyo-ku, Tokyo 113-8656, Japan.*
[2]*CREST, Japan Science and Technology Agency, 4-1-8 Honcho, Kawaguchi, Saitama 332-0012, Japan.*
[3]*Department of Electrical and Electronic Engineering, Tokyo Institute of Technology, 2-12-1 Ookayama, Meguro, Tokyo 152-8550, Japan.*
[4]*Department of Physics, Ho Chi Minh City University of Education, 280 An Duong Vuong Street, District 5, Ho Chi Minh City 738242, Vietnam.*
[5]*Institute of Engineering Innovation, The University of Tokyo, 7-3-1 Hongo, Bunkyo-ku, Tokyo 113-8656, Japan.*
[6]*PRESTO, Japan Science and Technology Agency, 4-1-8 Honcho, Kawaguchi, Saitama 332-0012, Japan*
[7]*Center for Spintronics Research Network (CSRN), The University of Tokyo, 7-3-1 Hongo, Bunkyo, Tokyo 113-8656, Japan.*

**Abstract**

Spin injection using ferromagnetic semiconductor at room temperature is a building block for the realization of spin-functional semiconductor devices. Nevertheless, this has been very challenging due to the lack of reliable room-temperature ferromagnetism in well-known group IV and III-V based semiconductors. Here, we demonstrate room-temperature spin injection by using spin pumping in a (Ga,Fe)Sb / BiSb heterostructure, where (Ga,Fe)Sb is a ferromagnetic semiconductor (FMS) with high Curie temperature ($T_C$) and BiSb is a topological insulator (TI). Despite the very small magnetization of (Ga,Fe)Sb at room temperature (45 emu/cc), we are able to detect spin injection from (Ga,Fe)Sb by utilizing the inverse spin Hall effect (ISHE) in the topological surface states of BiSb with a large inverse spin Hall angle of 2.5. Our study provides the first demonstration of spin injection as well as spin-to-charge conversion at room temperature in a FMS/TI heterostructure.

[#] goel@cryst.t.u-tokyo.ac.jp
[+] pham.n.ab@m.titech.ac.jp
[†] masaaki@ee.t.u-tokyo.ac.jp

## I.  Introduction

Room temperature injection and detection of spin-polarized carriers from ferromagnetic semiconductors (FMSs), which possess both ferromagnetic and semiconducting properties, are crucial for realizing future semiconductor-based spintronics devices [1],[2]. So far, successful spin injection has been reported only in prototypical Mn-doped III-V FMS (Ga,Mn)As at very low temperature < 120 K [3],[4],[5]. This is because these Mn-doped FMSs show ferromagnetic order only at low temperature (the maximum Curie temperature $T_C$ is 200 K in (Ga,Mn)As) [6]. On the other hand, Fe-doped III-V FMSs, such as (In,Fe)As, (Ga,Fe)Sb, and (In,Fe)Sb, are shown to be promising for spintronic device applications because of their high $T_C$ ($T_C$ > 300 K) [7]–[12]. Recently, we found that (Ga,Fe)Sb shows clear ferromagnetic resonance (FMR) at room temperature [9],[10], which is the basis for using FMSs as a spin injector in semiconductor spin devices operating at room temperature. So far, spin pumping [13],[14], electrical spin injection [15],[16], spin Seebeck effect [17], and other optical methods [18] have been used to study the spin injection from a ferromagnetic (FM) layer into a non-magnetic (NM) layer. Spin pumping, which employs FMR, is one of the most efficient methods, because a pure spin current can be transported through a FM/NM interface even under a conductivity mismatch condition [19]. The injected spin current is detected by the inverse spin Hall effect (ISHE) in the NM layer, which converts the spin current into a charge current via spin-orbit interaction (SOI) [20]. The efficiency of this spin-to-charge conversion is characterized by the inverse spin Hall angle (ISHA). Thus, to realize efficient ISHE, one needs spin Hall materials with strong SOI. Recently, it was found that the three-dimensional topological insulator (TI) BiSb has a strong SOI and large spin Hall angle (SHA) > 1 [21]. This large SHA was reported for both epitaxial and non-epitaxial BiSb layers [22], which means that poly-crystalline BiSb can also be used for spin detection by using its large SHA.

In this work, we prepare a BiSb/(Ga,Fe)Sb heterostructure, where we use (Ga,Fe)Sb ($T_C$ > 300 K) as the FMS layer and BiSb as the NM layer, and demonstrate spin injection by spin pumping and spin detection by ISHE at room temperature. Because the magnetization of (Ga,Fe)Sb is only 45 emu/cc at room temperature, which is 40 times smaller than that of Fe, detection of ISHE voltages is very challenging. To overcome this problem, we utilize the large SHA of the topological surface states of BiSb and detect ISHE voltages under FMR conditions. Furthermore, by carefully analyzing the voltage signals at various magnetic field directions, temperatures, and microwave powers, we distinguish the intrinsic ISHE signals from parasitic galvanomagnetic contributions, such as anomalous Hall effect (AHE) and planar Hall effect (PHE) [5],[23],[24]. We estimated the ISHA of poly-crystalline BiSb to be as large as 2.5. Our study provides the first step towards efficient spin injection and detection in practical semiconductor spintronics devices operating at room temperature.

## II. Sample growth and experimental procedure

Figure 1(a) shows the schematic structure of our sample. The growth of the sample structure is divided into two parts. We first grew heterostructures composed of $(Ga_{0.75},Fe_{0.25})Sb$ (50 nm)/AlSb (100 nm)/AlAs (10 nm)/GaAs (100 nm) on semi-insulating GaAs substrates with a growth rate of 0.5 μm/h by molecular beam epitaxy (MBE), followed by growth of 2 nm-thick GaSb cap layer to avoid surface oxidation. The substrate temperature ($T_S$) was 550 °C for the GaAs and AlAs layers, 470 °C for the AlSb layer, 250 °C for the (Ga,Fe)Sb and GaSb layer. After that, we preserved a half part of the sample as a reference sample and transferred the other half part of the sample to the sputtering system for the BiSb deposition. There, we first etched the GaSb cap layer using Ar-plasma at room temperature. Then, we deposited a 7 nm-thick $Bi_{0.85}Sb_{0.15}$ layer with a rate of 5.8 nm/min on top of the 50 nm-thick (Ga,Fe)Sb layer by co-sputtering Bi and Sb targets with Ar plasma at room temperature. The growth process was monitored *in situ* by reflection high-energy electron diffraction (RHEED). Figures 1(b)–(d) show RHEED patterns in the [$\bar{1}$10] direction of the sample structure. During the MBE growth, the (Ga,Fe)Sb thin film showed bright and streaky RHEED [Fig. 1(b)], thereby indicating good two-dimensional growth of a zinc-blende crystal structure. After etching the GaSb cap layer, we observed spotty RHEED, indicating that the (Ga,Fe)Sb surface is exposed [Fig. 1(c)]. Finally, we saw a ring RHEED pattern after depositing BiSb, indicating poly-crystalline BiSb in our sample [Fig. 1(d)]. In this way, we deposited a poly-crystalline BiSb film on top of the epitaxial (Ga,Fe)Sb thin film. We characterized the magnetic properties of the BiSb (7 nm)/(Ga,Fe)Sb (50 nm) heterostructure (sample A) and a reference (Ga,Fe)Sb (50 nm) thin film (sample B) using magnetic circular dichroism (MCD) spectroscopy and superconducting quantum interference device (SQUID) magnetometry (see Supplementary Material (SM) [25] for detailed characterizations). From the MCD and SQUID characterizations, we confirmed the intrinsic ferromagnetism of (Ga,Fe)Sb in both samples A and B, and that (Ga,Fe)Sb is not affected by the etching of the GaSb cap layer and the BiSb overgrowth. Also, we note that (Ga,Fe)Sb has in-plane magnetic anisotropy at room temperature but changes to perpendicular magnetic anisotropy at low temperatures [26]. These results are further supported by magnetic-field angle dependence of the FMR spectra of (Ga,Fe)Sb in samples A and B, which indicates that there is almost no difference in magnetic anisotropy and the FMR occurs at the same resonance fields in both samples (see SM [25] for more detail).

For spin pumping measurements, we used an electron spin resonance spectrometer whose cavity resonates in the transverse electric TE$_{011}$ mode at a microwave frequency of 9.1 GHz. Figure 1(e) shows the BiSb/(Ga,Fe)Sb bilayer structure and coordinate axes used in our spin pumping experiment. For electrical measurements, we have connected the two gold wires at both edges of the sample with a distance of $l = 2$ mm apart as shown in Fig. 1(e). For the measurements, a static magnetic field $\mu_0 H$ is applied along the [1$\bar{1}$0] direction in the film plane (i.e. $\theta_H = 90°$), which corresponds to the easy magnetization axis of (Ga,Fe)Sb. Here, $\theta_H$ is the out-of-plane angle between $H$ and the [001] axis. A microwave magnetic field $h$ was applied along the [110] axis. The detailed measurement procedure is described in SM [25].

### III. Results and discussion

Figure 2(a) shows the FMR spectra at various $\theta_H$ measured at a microwave power of 200 mW. The resonance field $\mu_0 H_R$ of the FMR spectra changes from 290 mT to 320 mT when $\theta_H$ is changed from ±90° ($H$ // [1$\bar{1}$0] and [$\bar{1}$10]) to 0° ($H$ // [001]). This indicates that the 50 nm-thick (Ga$_{0.75}$,Fe$_{0.25}$)Sb film in this work has the in-plane magnetic anisotropy with an easy magnetization axis along the [1$\bar{1}$0] axis at 300 K. Figure 2(b) shows the electromotive force (EMF) signal $V$ (offset voltage $V_{offset}$ is subtracted) at various $\theta_H$. As shown in Fig. 2(b), the $V - H$ curves exhibit voltage extrema corresponding to the resonance fields ($\mu_0 H_R$) in the FMR spectra, which indicates that the observed voltage signals were generated by FMR. Also, the $\theta_H$ - dependence of EMF is consistent with the formula EMF $\propto j_S \times \sigma$, where $j_S$ and $\sigma$ are the spin current and spin polarization vector of the spin current, respectively [27],[28]. When $\theta_H$ is changed from 90° to –90°, the voltage signal changes its sign from positive to negative, respectively. The maximum EMF signal is obtained at $\theta_H = 90°$ and –90°, where $j_S$ is perpendicular to the $\sigma$. On the other hand, the EMF signal disappears at $\theta_H = 0°$, when $j_S$ is parallel to $\sigma$. This indicates that the observed EMF signal is consistent with the well-established model of spin injection by spin pumping and spin-to-charge conversion by ISHE.

Next, we investigated the microwave power dependence of EMF to further confirm the ISHE in our sample. Figures 3(a) and 3(b) show the microwave power dependence of EMF for sample A at 300 K for $\theta_H = 90°$ ($H$ // $[1\bar{1}0]$) and $-90°$ ($H$ // $[\bar{1}10]$), respectively. We note that in both directions, the EMF is proportional to the applied microwave power, which discards the contribution of the Seebeck effect [29]. Next, for qualitative analysis, we decomposed the obtained EMF into a symmetric (Lorentzian) voltage component $V_{sym}$ and an asymmetric (dispersive) voltage component $V_{asym}$, using the following fitting expression:

$$\text{EMF}\ (V - V_{\text{offset}}) = V_{sym} \frac{(\mu_0 \Delta H)^2}{(\mu_0 H - \mu_0 H_R)^2 + (\mu_0 \Delta H)^2} + V_{asym} \frac{-2\mu_0 \Delta H (\mu_0 H - \mu_0 H_R)}{(\mu_0 H - \mu_0 H_R)^2 + (\mu_0 \Delta H)^2}, \quad (1)$$

where $\mu_0 \Delta H$ is the half-width at half maximum of the FMR linewidth and $\mu_0 H_R$ is the resonance field. Here, $V_{sym}$ includes contributions from the ISHE and PHE, while $V_{asym}$ originates from the AHE [5]. As shown in Fig. 3(a) and 3(b), the experimental data (solid curves) is well reproduced by the fitting function (black dotted curves) of Eq. (1). From Fig. 3(c), the estimated $V_{sym}$ and $V_{asym}$ components are directly proportional to the applied microwave power, which is an important identity of ISHE. This is also evidenced by the $\theta_H$ dependence of the EMF [Fig. 2(b)], where we observed nearly similar voltage extrema ~6 µV when = 90° ($H$ // $[1\bar{1}0]$) and -90°($H$ // $[\bar{1}10]$).

Next, in order to distinguish the contribution of spin-pumping to $V_{sym}$ from that of PHE in sample A [(Ga,Fe)Sb/BiSb], we performed the FMR measurements on the reference sample B [(Ga,Fe)Sb without BiSb]. In Figs. 4 (a)–(c), we compare the EMF signals of sample A and sample

B at various $\theta_H$. While we notice both the strong symmetric and asymmetric components in sample A, we observe only the asymmetric component in sample B. This result suggests that the contribution of the PHE to $V_{sym}$ in (Ga,Fe)Sb is negligible. We then use Eq. (1) to fit to the voltage signals and estimate the voltage components $V_{sym}$ and $V_{asym}$ for sample A and $V_{sym}^{ref}$ and $V_{asym}^{ref}$ for reference sample B. Figure 4 (d) shows the estimated voltage components as a function of $\theta_H$. We see that $V_{sym}^{ref} \ll V_{asym}^{ref}$ (~100 times different) for sample B, which means that the voltage contribution from PHE of (Ga,Fe)Sb is negligible at 300 K. Meanwhile, the contribution from AHE of (Ga,Fe)Sb at 300 K to the asymmetric voltage component exists in both samples. However, in the case of sample A, $V_{sym} > V_{asym}$, which means that ISHE is stronger and dominant over AHE.

Next, we investigated the temperature dependence of the spin injection by spin pumping from (Ga,Fe)Sb to BiSb. First, we measure the resistivity of both samples with a Hall bar size of 1 mm × 0.5 mm using the 4-terminal method. Figure 5(a) shows the temperature dependence of the resistivity of samples A and B. We note that sample A (with the BiSb layer, red circles) shows larger conductivity than sample B (without BiSb, black circles) at all temperatures, which reflects the much higher electrical conductivity of the BiSb layer [30]. The estimated conductivities ($\sigma_{GaFeSb}$, $\sigma_{BiSb}$) of (Ga,Fe)Sb and BiSb at 300 K are shown in Table I. Since $\sigma_{BiSb}$ (1.8×10$^5$ $\Omega^{-1}$m$^{-1}$) $\gg \sigma_{GaFeSb}$ (1.4×10$^3$ $\Omega^{-1}$m$^{-1}$), the charge current mostly flows in the BiSb layer. In Ref. 22 and Ref. 30, it was reported that a 10 nm-thick BiSb layer has a large bandgap of ~200 meV due to the quantum size effect, and that the topological surface states (SSs) account for 97% of the total current in BiSb even at room temperature. Because we used the 7 nm-thick BiSb in this study, it is likely that nearly 100% of the current flows on SSs of BiSb. We will use this information later for the estimation of the ISHA of BiSb.

We then evaluated the temperature dependence of AHE in reference sample B. For this, we prepared a Hall bar of 200 μm × 50 μm and measure the temperature dependence of AHE with $H$ // [001]. Figure 5(b) shows the anomalous Hall resistance ($R_{AHE}$) as a function of temperature, which increases with decreasing temperature. We know that the saturation values of $R_{AHE}$, $R_{AHE}^0$, reflects the AHE from the (Ga,Fe)Sb film; however, due to the presence of the ordinary Hall effect (OHE) along with the AHE, it is difficult to saturate $R_{AHE}$, especially at high temperatures. Thus, we use $R_{AHE}$ values near 1T as $R_{AHE}^0$ which is closer to the saturation values of $R_{AHE}$. Figure 5(c) shows $R_{AHE}^0$ as a function of temperature in sample B. In Fig. 5(c), we see a clear increase in $R_{AHE}^0$ with decreasing temperature. We also measured the planar Hall resistance ($R_{PHE}$) as a function of in-plane $H$ angle $\varphi_H$ (where $\varphi_H$ is the angle between $H$ and the [1$\bar{1}$0] axis), at a constant $\mu_0 H$ of 50 mT, 80 mT, and 100 mT at 300 K [Fig. 5(d)]. We note that in comparison with $R_{AHE}^0$ (= ~9 Ω), $R_{PHE}$ (= ~0.6 Ω) is much smaller at 300 K. Furthermore, although $R_{PHE}$ would follow $\cos(2\varphi_H)$ dependence, $R_{PHE}$ is actually dominated by the $\cos(\varphi_H)$ component [see Fig. 5(d)], indicating that the AHE contribution is dominant. This is probably because the in-plane magnetic field is misaligned and has a perpendicular component, leading to the AHE contribution, while the PHE contribution of $R_{PHE}^0 \cos(2\varphi_H)$ is negligible in Fig. 5(d), where $R_{PHE}^0$ is the magnitude of $R_{PHE}$. Therefore, we conclude $R_{PHE}^0 \ll R_{AHE}^0$ at room temperature. This result is consistent with the absence of the $V_{sym}$ component in sample B observed in Fig. 4(d).

Finally, we measured the temperature dependence of the EMF at $\theta_H = 90°$ in sample A, as shown in Fig. 5(e). We found that the EMF peak position appears at nearly the same resonance field ($\mu_0 H_R$

= 290 mT) at all temperatures. However, the magnitude and shape of the EMF signal depend on temperature, suggesting the temperature dependent contribution of other galvanomagnetic effects, especially at low temperature, to our EMF signals. For example, the EMF peak at 10 K is significantly broader than that at 300 K. We then fit Eq. (1) to the experimental data to estimate $V_{sym}$ and $V_{asym}$ at each temperature. Figure 5(f) shows the temperature dependence of $V_{sym}$ and $V_{asym}$. At low temperatures, we see clear enhancement of $V_{asym}$ by a factor of 3 at 10 K as compared with that at 300 K, which is consistent with the temperature dependence of $R_{AHE}$ shown in the Fig. 5(c). Meanwhile, $V_{sym}$ shows close values (4.1 – 5.2 µV) at all temperatures, suggesting that the contribution of PHE to $V_{sym}$ is also negligible at low temperatures. Furthermore, it was shown that the emergence of PHE could lead to the shift of the EMF peak position at low temperature [24]; nevertheless, we observed no such change in Fig. 5(e). Therefore, we conclude that the contribution of PHE in our sample is negligible.

Next, we estimated ISHA by using the well-known expression:

$$V_{ISHE} = \frac{l\theta_{ISHE}\lambda_{BiSb}\tanh(d_{BiSb}/2\lambda_{BiSb})}{\sigma_{GaFeSb}d_{GaFeSb} + \sigma_{BiSb}d_{BiSb}}\left(\frac{2e}{\hbar}\right)j_S^{BiSb/GaFeSb}, \qquad (2)$$

where, $\sigma_{GaFeSb}$, $\sigma_{BiSb}$, $d_{GaFeSb}$, $d_{BiSb}$ are the conductivities and thicknesses of the (Ga,Fe)Sb and BiSb layers, respectively; $l$ is the length between the electrodes, $e$ is the electron charge, and $\hbar$ is the Dirac constant; $\theta_{ISHE}$ is the inverse spin Hall angle; $j_S^{BiSb/GaFeSb}$ is the spin current density at the interface (see SM [25] for $j_S^{BiSb/GaFeSb}$). The estimated values of these parameters are shown in Table I. As we discussed above, conduction of the charge current in BiSb occurs only in the SSs of the BiSb. Thereby, we modify Eq. (2) by replacing $d_{BiSb}$ by $d_S$ and $\lambda_{BiSb}$ by $\lambda_S$ in the numerator part, where $d_S$ and $\lambda_S$ are the thickness and spin diffusion length of the SSs in the BiSb layer. We further assume that the spin-diffusion length of SSs in $Bi_{0.85}Sb_{0.15}$ is the same as that in Bi (∼1 nm) [31],[32]. The modified expression for ISHA estimation based on the dominant surface conduction is given by

$$V_{\text{ISHE}} = \frac{l\theta_{\text{ISHE}} d_S \tanh(1/2)}{\sigma_{\text{BiSb}} d_{\text{BiSb}}} \left(\frac{2e}{\hbar}\right) j_S^{\text{BiSb/GaFeSb}}, \quad (3)$$

Using the parameters in Table I, Eq. (3), and $d_S \sim \lambda_S \sim 1$ nm, we roughly estimated the effective ISHA value (mainly from the SSs conduction) to be $\theta_{\text{ISHA}}^{\text{eff}} \sim 2.5$. This $\theta_{\text{ISHA}}^{\text{eff}}$ is close to the SHA value ($\theta_{\text{SHA}} \sim 3.2$) reported for non-epitaxial BiSb [22], and is larger than that for GaAs, Pt, Si, $Bi_2Se_3$, and other heavy elements [5],[33],[34],[35]. In the present study, we deposited BiSb by sputtering on MBE-grown (Ga,Fe)Sb, which may have led to lower interface quality. Thus, an even higher ISHA value may be obtained by growing both BiSb and (Ga,Fe)Sb epitaxially, which is yet to be achieved in the future. Furthermore, the result presented in this study on Fe-doped FMS (Ga,Fe)Sb is clearly different from that of prototypical Mn-doped FMS (Ga,Mn)As in the following ways: (i) While the symmetric voltage signal component has a strong contribution from PHE (~88%) in (Ga,Mn)As, (Ga,Fe)Sb shows negligible PHE. (ii) While spin-to-charge conversion was observed at low temperature (< 120 K) in (Ga,Mn)As [5] due to its low $T_C$, (Ga,Fe)Sb shows spin pumping signals at 300 K and potentially even at higher temperature because of its high $T_C$.

## IV. Conclusion

We have successfully demonstrated spin injection by spin pumping and detection by ISHE in a BiSb/(Ga,Fe)Sb heterostructure at room temperature. This work is the first spin pumping experiment using high-$T_C$ FMS and high-performance TI. From the temperature, microwave power, and magnetic field direction dependences of the ISHE voltage as well as the spin pumping experiment on the reference (Ga,Fe)Sb sample, we conclude that the symmetric voltage component obtained in the BiSb/(Ga,Fe)Sb sample has a negligible galvanomagnetic effect. We have estimated the ISHA of about 2.5 for the BiSb layer, which is comparable to its SHA value reported previously. The result presented in this study opens new opportunities for semiconductor spintronic device applications operating at room temperature.

**ACKNOWLEDGMENTS**

This work was partly supported by Grants-in-Aid for Scientific Research (No. 20H05650, No. 18H03860, and No. 17H04922), CREST of JST (JPMJCR1777, JPMJCR18T5), the Spintronics Research Network of Japan (Spin-RNJ), and the Murata Science Foundation. A part of this work was conducted at the Advanced Characterization Nanotechnology Platform of the University of Tokyo, supported by the 'Nanotechnology Platform' of the Ministry of Education, Culture, Sports, Science, and Technology (MEXT), Japan. N.H.D.K. acknowledges Marubun Research Promotion Foundation for an exchange research grant and JSPS for a postdoctoral fellowship for research in Japan (P20050).

**REFERENCES**

[1]  H. Ohno, Science **281**, 951–956 (1998).

[2]  M. Oestreich, Nature **402**, 735 (1999).

[3]  Y. Chye, M. E. White, E. Johnston-Halperin, B. D. Gerardot, D. D. Awschalom, and P. M. Petroff, Phys. Rev. B **66**, 201301, (2002).

[4]  M. Kohda, Y. Ohno, F. Matsukura, H. Ohno, Physica E Low Dimens. Syst. Nanostruct. **32**, 438 (2006).

[5]  L. Chen, F. Matsukura, and H. Ohno, Nat. Commun. **4**, 2055 (2013).

[6]  L. Chen, X. Yang, F. Yang, J. Zhao, J. Misuraca, P. Xiong, and S. Von Molnár, Nano Lett. **11**, 2584 (2011).

[7]  P. N. Hai, M. Yoshida, A. Nagamine, and M. Tanaka, Jpn. J. Appl. Phys. **59** 063002 (2020).

[8]  N. T. Tu, P. N. Hai, L. D. Anh, and M. Tanaka, Appl. Phys. Lett. **108**, 192401 (2016).

[9]  S. Goel, L. D. Anh, S. Ohya, and M. Tanaka, Phys. Rev. B **99**, 014431 (2019).

[10] S. Goel, L. D. Anh, N. T. Tu, S. Ohya, and M. Tanaka, Phys. Rev. Materials **3**, 084417 (2019).

[11] N. T. Tu, P. N. Hai, L. D. Anh, and M. Tanaka, Appl. Phys. Express **11**, 063005 (2018).

[12] N. T. Tu, P. N. Hai, L. D. Anh, and M. Tanaka, Appl. Phys. Express **12** 103004 (2019).

[13] S. Mizukami, K. Ando, and T. Miyazaki, J. Magn. Magn. Mater. **239**, 42 (2002).

[14] K. Ando and E. Saitoh, J. Appl. Phys. **108**, 113925 (2010).

[15] M. Johnson and R. H. Silsbee, Phys. Rev. Lett. **55**, 1790–1793 (1985)

[16] Y. Ohno, D. K. Young, B. Beschoten, F. Matsukura, H. Ohno, and D. D. Awschalom, Nature **402**, 790–792 (1999).

[17] K. Uchida, S. Takahashi, K. Harii, J. Ieda, W. Koshibae, K. Ando, S. Maekawa, and E. Saitoh, Nature **455**, 778 (2008).

[18] R. Fiederling, M. Keim, G. Reuscher, W. Ossau, G. Schmidt, A. Waag, and L. W. Molenkamp, Nature **402**, 787–790 (1999).

[19] K. Ando, S. Takahashi, J. Ieda, H. Kurebayashi, T. Trypiniotis, C. H. W. Barnes, S. Maekawa, and E. Saitoh, Nature Mater. **10**, 655–659 (2011).

[20] E. Saitoh, M. Ueda, H. Miyajima, and G. Tatara, Appl. Phys. Lett. **88**, 182509 (2006).

[21] N. H. D. Khang, Y. Ueda, and P. N. Hai, Nature Mater. **17**, 808–813 (2018).

[22] N. H. D Khang, S. Nakano, T. Shirokura, Y. Miyamoto, and P. N. Hai, Scientific Reports, **10**, 12185 (2020).

[23] H. J. Juretschke, J. Appl. Phys. **31**, 1401 (1960).

[24] S. Ohya, A. Yamamoto, T. Yamaguchi, R. Ishikawa, R. Akiyama, L. D. Anh, S. Goel, Y. K. Wakabayashi, S. Kuroda, and M. Tanaka, Phys. Rev. B **96**, 094424 (2017).

[25] Supplementary Material.

[26] S. Goel, L. D. Anh, S. Ohya, and M. Tanaka, J. Appl. Phys. **127**, 023904 (2020).

[27] A. Azevedo, L. H. Vilela Leão, R. L. Rodriguez-Suarez, A. B. Oliveira, and S. M. Rezende, J. Appl. Phys. **97**, 10C715 (2005).

[28] K. Ando, S. Takahashi, J. Ieda, Y. Kajiwara, H. Nakayama, T. Yoshino, K. Harii, Y. Fujikawa, M. Matsuo, S. Maekawa, and E. Saitoh J. Appl. Phys. **109**, 103913 (2011).

[29] Y. Shiomi, K. Nomura, Y. Kajiwara, K. Eto, M. Novak, K. Segawa, Y. Ando, and E. Saitoh, Phys. Rev. Lett. **113**, 196601 (2014).

[30] T. Fan, M. Tobah, T. Shirokura, N. H. D. Khang, and P. N. Hai, Jpn. J. Appl. Phys. **59**, 063001 (2020).

[31] T. Hirahara, T. Nagao, I. Matsuda, G. Bihlmayer, E. V. Chulkov, Yu. M. Koroteev, P. M. Echenique, M. Saito, and S. Hasegawa, Phys. Rev. Lett. **97**, 146803 (2006).

[32] D. Hou, Z. Qiu, K. Harii, Y. Kajiwara, K. Uchida, Y. Fujikawa, H. Nakayama, T. Yoshino, T. An, K. Ando, X. Jin, and E. Saitoh, Appl. Phys. Lett. **101**, 042403 (2012).

[33] J.-C. Rojas-Sánchez, N. Reyren, P. Laczkowski, W. Savero, J.-P. Attané, C. Deranlot, M. Jamet, J.-M. George, L. Vila, and H. Jaffrès, Phys. Rev. Lett. **112**, 106602 (2014).

[34] K. Ando and E. Saitoh, Nat. Commun. **3**, 629 (2012).

[35] P. Deorani, J. Son, K. Banerjee, N. Koirala, M. Brahlek, S. Oh, and H. Yang, Phys. Rev. B **90**, 094403 (2014).

TABLE I. Parameters used for the ISHA estimation. $l$ is the distance between the gold wire contacts, $d_{\text{GaFeSb}}$ and $d_{\text{BiSb}}$ are the thickness of the (Ga,Fe)Sb layer and the BiSb layer, respectively. $\sigma_{\text{GaFeSb}}$ and $\sigma_{\text{BiSb}}$ are the electrical conductivity estimated from transport measurements, respectively. $j_S^{\text{BiSb/GaFeSb}}$ is the spin current density at the BiSb/GaFeSb interface estimated by the spin pumping measurement. $V_{\text{ISHE}}$ (= $V_{\text{sym}}$) is the ISHE voltage signal estimated as the maximum peak height of the symmetric component of the voltage signal after removing the offset value. All parameters are estimated at 300 K, at a microwave power of 200 mW with a magnetic field $H$ direction of $\theta_H = 90°$ ($H$ // [1$\bar{1}$0]).

| $l$ (mm) | $d_{\text{GaFeSb}}$ (nm) | $d_{\text{BiSb}}$ (nm) | $\sigma_{\text{GaFeSb}}$ ($\Omega^{-1}$ m$^{-1}$) | $\sigma_{\text{BiSb}}$ ($\Omega^{-1}$ m$^{-1}$) | $j_S^{\text{BiSb/GaFeSb}}$ (J m$^{-2}$) | $V_{\text{ISHE}}$ (μV) |
|---|---|---|---|---|---|---|
| 2 | 50 | 7 | $1.4 \times 10^3$ | $1.8 \times 10^5$ | $5.7 \times 10^{-12}$ | 4.4 |

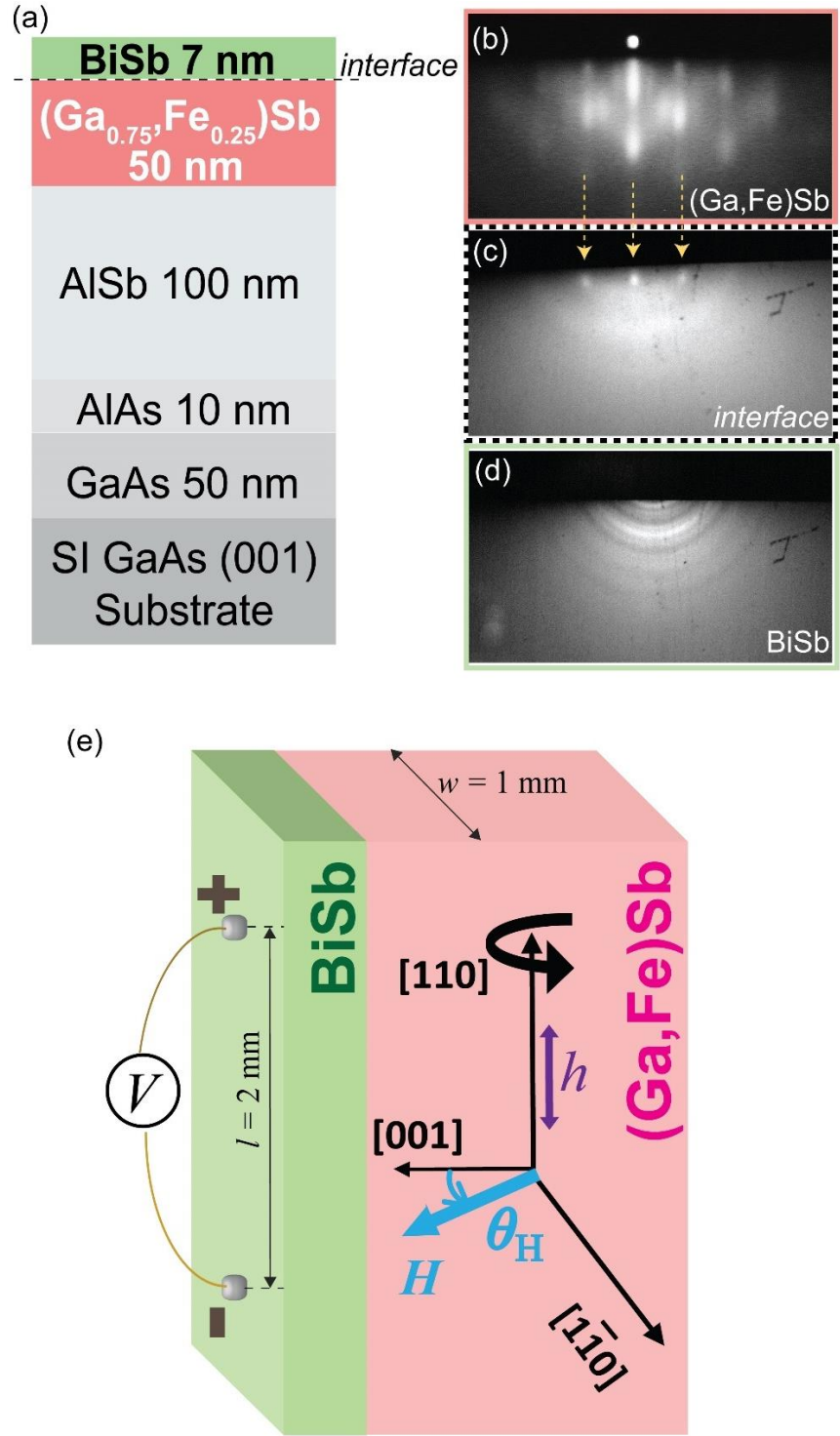

FIG. 1. (a) Schematic illustration of the (001)-oriented sample structure composed of BiSb (7 nm)/ (Ga,Fe)Sb (50 nm)/ AlSb (100 nm)/ AlAs (10 nm)/ GaAs (50 nm) grown on a semi-insulating GaAs(001) substrate. (b)–(d) *In-situ* reflection high energy electron diffraction (RHEED) patterns

observed along the [1̄10] axis with streaky RHEED during the MBE growth of (Ga,Fe)Sb (b), spotty RHEED of the interface after anti-sputtering (c), and circular rings RHEED of BiSb during sputtering (d). The black dotted line in (c) corresponds to the interface between BiSb and (Ga,Fe)Sb layer. (e) Sample alignment and coordinate system used in the spin pumping measurement. A microwave magnetic field $h$ was applied along the [110] axis of the sample. $\theta_H$ is the angle of the magnetic field $H$ with respect to the [001] axis, respectively, where $H$ is in the (110) plane.

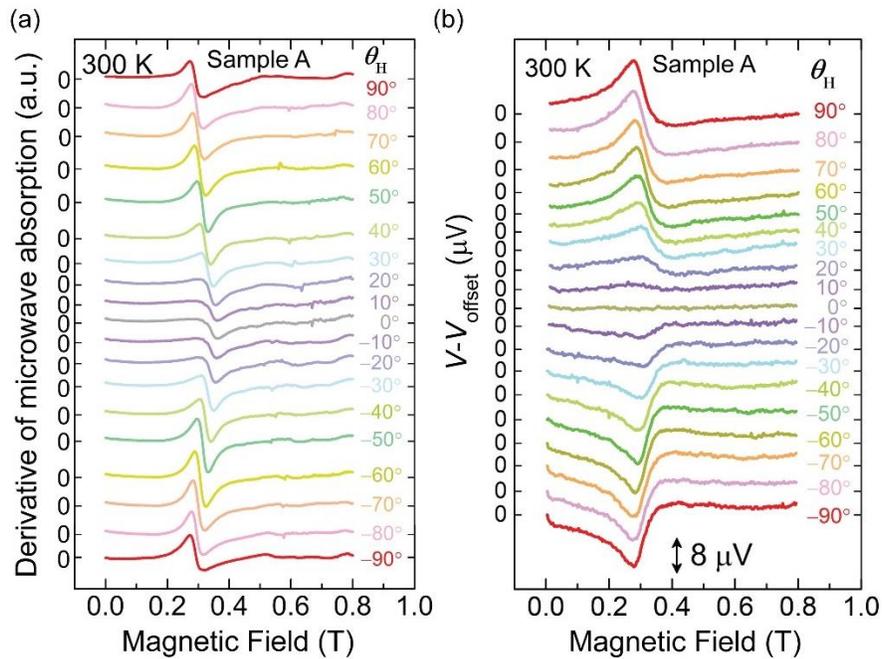

FIG. 2. (a) Ferromagnetic resonance (FMR) spectra and (b) corresponding electromotive force (EMF) voltage peaks measured at various magnetic field $H$ directions ($\theta_H$) at 300 K.

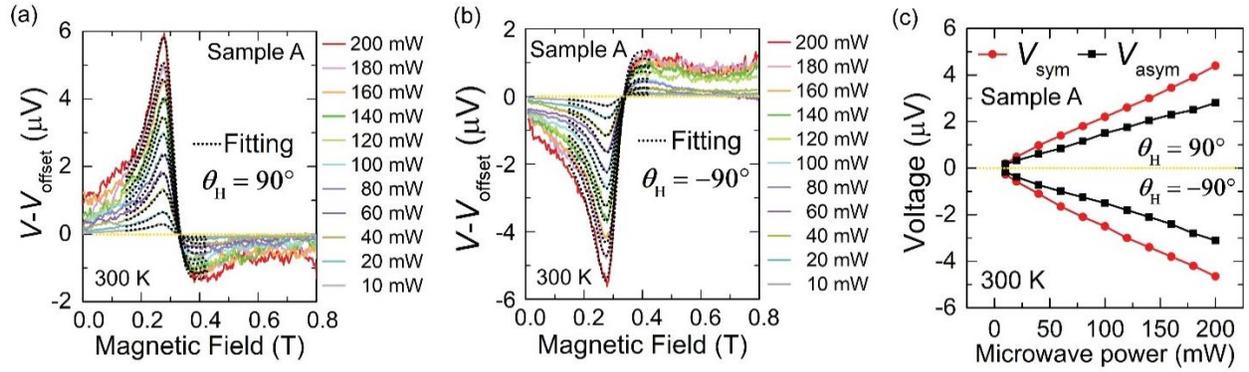

FIG. 3. (a) and (b) Magnetic field $H$ dependences of the voltage signal $V$ (offset voltage $V_{\text{offset}}$ is subtracted) in sample A measured at 300 K at various microwave powers, ranging from 10 mW to 200 mW, for $\theta_H = 90°$ and $-90°$, respectively. The black dotted curves represent fitting which includes a summation of the symmetric and antisymmetric voltage components. (c) Microwave power dependence of the symmetric voltage component $V_{\text{sym}}$ and antisymmetric voltage component $V_{\text{asym}}$ at $\theta_H = 90°$ and $-90°$.

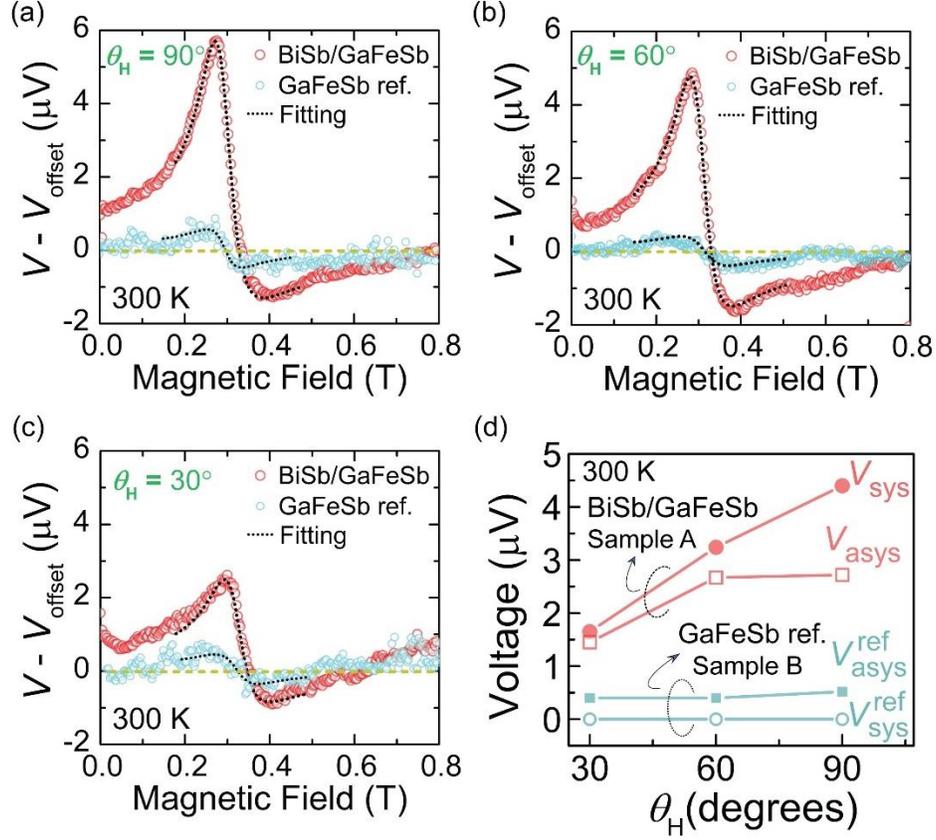

FIG. 4. (a)–(c) Voltage signal $V - V_{\text{offset}}$ for the BiSb/(Ga,Fe)Sb heterostructure (sample A, red big circles) and reference (Ga,Fe)Sb sample (sample B, green small circles) observed for various magnetic field $\boldsymbol{H}$ directions ($\theta_H$) at 300 K, where $\boldsymbol{H}$ is in the (110) plane. The microwave power is 200 mW. (d) The derived symmetric voltage component ($V_{\text{sym}}$) and antisymmetric voltage component ($V_{\text{asym}}$) of both the samples at 300 K. The circles and squares represent $V_{\text{sym}}$ and $V_{\text{asym}}$, respectively.

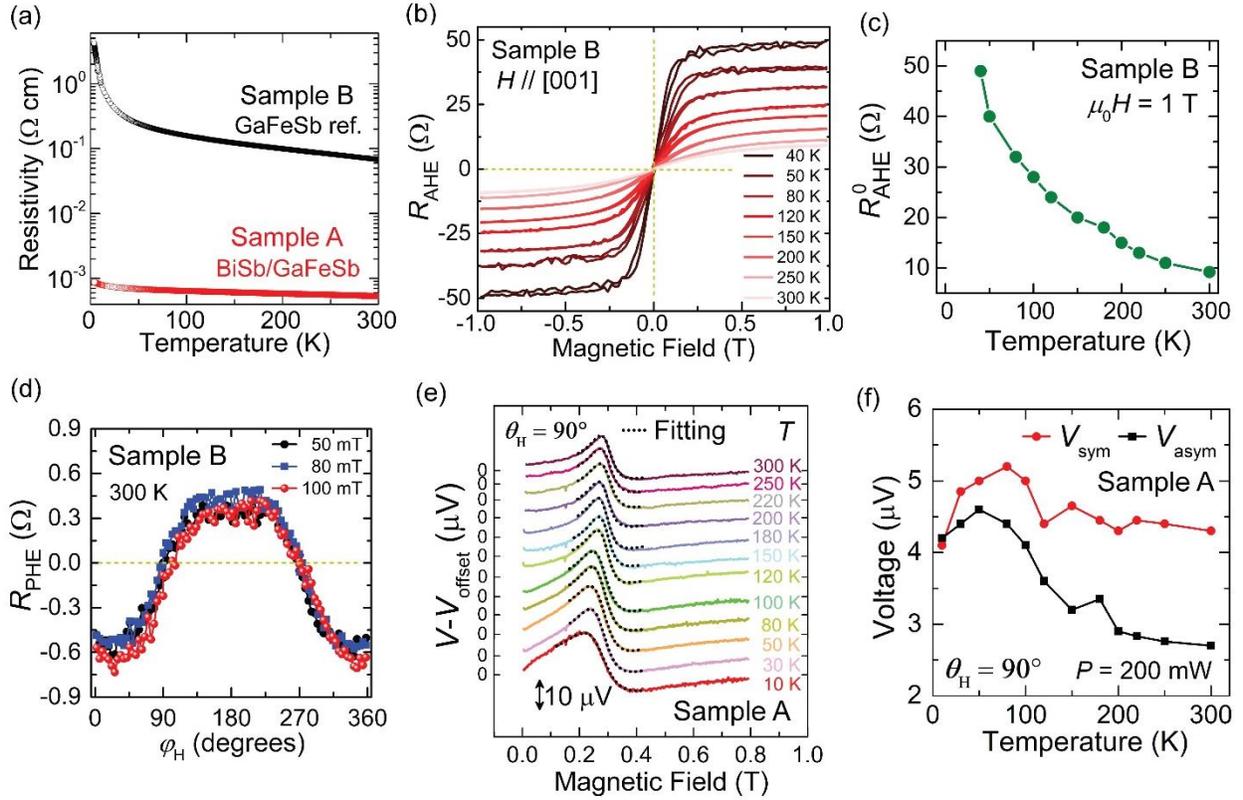

FIG. 5. (a) Temperature dependence of the resistivity of the BiSb (7 nm) / (Ga,Fe)Sb (50 nm) heterostructure (sample A, red open circles) and the 50 nm-thick (Ga,Fe)Sb reference (sample B, black open circles). (b) Perpendicular magnetic field $H$ dependence of the anomalous Hall resistance ($R_{AHE}$) of sample B measured at various temperatures. (c) Saturation values ($R_{AHE}^0$) of $R_{AHE}$ as a function of temperature in sample B. (d) In-plane magnetic field direction ($\varphi_H$) dependence of the planar Hall resistance ($R_{PHE}$) of sample B with the applied magnetic field of 50 mT, 80 mT, and 100 mT measured at 300 K. $\varphi_H$ is the angle of the in-plane magnetic field with respect to the [1$\bar{1}$0] axis. (e) $H$ dependences of the voltage signal $V$ (offset voltage $V_{offset}$ is subtracted) of sample A measured with 200 mW at various temperatures, ranging from 300 K to 10 K, for $\theta_H = 90°$. The black dotted curves represent fittings which were used to derive the symmetric voltage ($V_{sym}$) and antisymmetric voltage ($V_{asym}$) components. (f) $V_{sym}$ and $V_{asym}$ of sample A as a function of temperature. Red solid circles and black solid squares represent $V_{sym}$ and $V_{asym}$, respectively.

# Supplementary Material
# Room-temperature spin injection and spin-to-charge conversion in a ferromagnetic semiconductor / topological insulator heterostructure

Shobhit Goel,[1,2,#] Nguyen Huynh Duy Khang,[3,4] Le Duc Anh,[1,5,6] Pham Nam Hai,[2,3,7,+] and Masaaki Tanaka[1,2,7,†]

[1]*Department of Electrical Engineering and Information Systems, The University of Tokyo, 7-3-1 Hongo, Bunkyo-ku, Tokyo 113-8656, Japan.*
[2]*CREST, Japan Science and Technology Agency, 4-1-8 Honcho, Kawaguchi, Saitama 332-0012, Japan.*
[3]*Department of Electrical and Electronic Engineering, Tokyo Institute of Technology, 2-12-1 Ookayama, Meguro, Tokyo 152-8550, Japan.*
[4]*Department of Physics, Ho Chi Minh City University of Education, 280 An Duong Vuong Street, District 5, Ho Chi Minh City 738242, Vietnam.*
[5]*Institute of Engineering Innovation, The University of Tokyo, 7-3-1 Hongo, Bunkyo-ku, Tokyo 113-8656, Japan.*
[6]*PRESTO, Japan Science and Technology Agency, 4-1-8 Honcho, Kawaguchi, Saitama 332-0012, Japan*
[7]*Center for Spintronics Research Network (CSRN), The University of Tokyo, 7-3-1 Hongo, Bunkyo, Tokyo 113-8656, Japan.*

## 1. Magnetic circular dichroism (MCD) measurements

We characterized the intrinsic ferromagnetism of (Ga,Fe)Sb before and after depositing BiSb by sputtering, using the magnetic circular dichroism (MCD) spectroscopy in a reflection setup. The reflection MCD intensity is given by the difference in the optical reflectivity for right ($R_{\sigma+}$) and left ($R_{\sigma-}$) circular polarization of light. The MCD intensity is expressed as $MCD = \frac{90}{\pi} \frac{(R_{\sigma+} - R_{\sigma-})}{2R} = \frac{90}{\pi} \frac{1}{2R} \frac{dR}{dE} \Delta E$, where $R$ is the optical reflectivity, $E$ is the photon energy and $\Delta E$ is the Zeeman splitting energy. Since MCD is proportional to $dR/dE$ and $\Delta E$, it directly probes the spin-polarized band structure of the measured material. Thus, MCD is a powerful tool to characterize the intrinsic magnetic properties of (Ga,Fe)Sb. Figures S1 (a) and (b) show the MCD spectra of sample A [BiSb (7 nm) / (Ga,Fe)Sb (50 nm)] and sample B [(Ga,Fe)Sb (50 nm)], respectively, measured at 5 K with a perpendicular magnetic field of 0.2 T, 0.5 T, and 1 T, normalized by the intensity of the $E_1$ peak at 1 T. As shown in Figs. S1(a) and (b), the MCD spectra have the $E_1$ peak around ~ 2.4 eV, reflecting the band structure of zinc-blende (Ga,Fe)Sb in both samples. In Figs. S1(a) and (b), normalized MCD spectra which were measured at various magnetic fields are overlapped on one spectrum, indicating that the ferromagnetism in (Ga,Fe)Sb comes from a single phase, that is the zinc-blende (Ga,Fe)Sb. Also, Figs. S1(c) and (d) show the magnetic-field ($\mu_0 H$) dependence of the MCD intensity at $E_1$ of both the samples, which show clear hysteresis even at room temperature. Thus, the ferromagnetism and the related magnetic properties are maintained in the (Ga,Fe)Sb before and after depositing BiSb without destroying the intrinsic ferromagnetism of zinc-blende (Ga,Fe)Sb. Figures S1(e) and (f) show the corresponding Arrott plots, which indicate that the Curie temperature $T_C$ is higher than 320 K in both samples. We estimated $T_C$ by extrapolating the Arrott plots measured at 290 – 320 K. As shown in insets of Figs. S1 (e) and (f), we found that $T_C$ is slightly enhanced from 345 K in sample B to 355 K in sample A; this may be caused by a magnetic proximity effect and yet to be studied in detail in our future work.

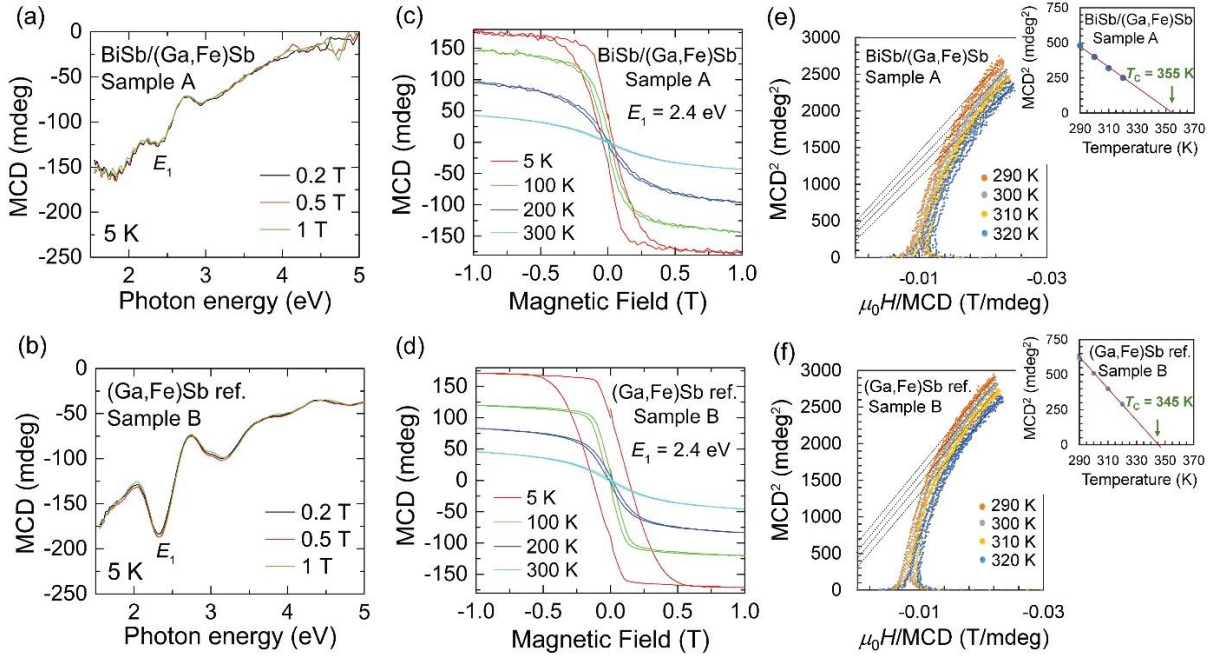

figure S1. (a) and (b) MCD spectra of the (a) BiSb (7nm) / (Ga,Fe)Sb (50 nm) heterostructure (sample A) and (b) (Ga,Fe)Sb (50 nm) reference (sample B) at 5 K with a magnetic field of 0.2 T, 0.5 T, and 1 T applied perpendicular to the film plane. The MCD spectra measured at 0.2 T and 0.5 T are normalized to that at 1 T by the intensity of the $E_1$ peak. (c) and (d) $MCD - H$ curves of (c) BiSb/(Ga,Fe)Sb (sample A) and (d) (Ga,Fe)Sb reference (sample B) at various temperatures. (e) and (f) Arrott plots $MCD^2 - H/MCD$ of the (e) BiSb/(Ga,Fe)Sb (sample A) and (f) (Ga,Fe)Sb (sample B). Insets show the Curie temperature ($T_C$) estimated by extrapolating the Arrott plots measured at 290 – 320 K of both the samples. $T_C$ is slightly enhanced from 345 K in sample B to 355 K in sample A.

## 2. Superconducting quantum interference device (SQUID) magnetometry measurements

Figures S2(a) and (b) show the magnetic-field dependence of the magnetization ($M - H$ characteristics) of sample A [BiSb/(Ga,Fe)Sb] and sample B [(Ga,Fe)Sb] measured at 10 K, with a magnetic field $H$ applied along the in-plane [110] axis (red solid circles) and the perpendicular [001] axis (black open circles). In both samples, the saturation magnetization is nearly the same in both samples, indicating that the ferromagnetism of (Ga,Fe)Sb did not change before and after depositing BiSb. *M* saturates at smaller *H* when ***H*** // [001] than when ***H*** // [110] indicating perpendicular magnetic anisotropy at 10 K. Note that (Ga,Fe)Sb has in-plane magnetic anisotropy at room temperature (as shown in the FMR result of Fig. S3 or Fig.2 in the main manuscript), but it changes to perpendicular magnetic anisotropy at low temperatures, as shown in Fig. S2 [ref. 26 in the main manuscript].

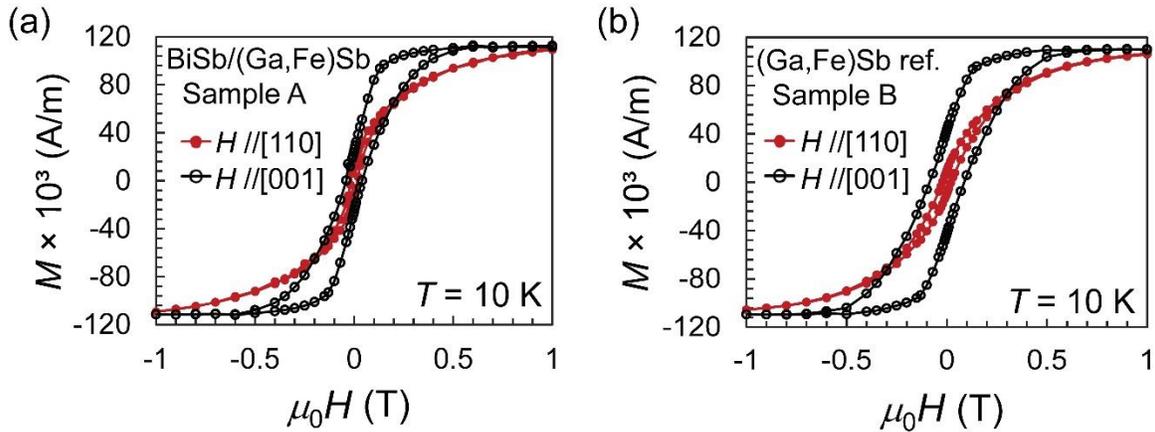

Figure S2. (a) and (b) Magnetization hysteresis curves (*M* – *H*) measured at 10 K for the (a) BiSb/(Ga,Fe)Sb (sample A) and (b) reference (Ga,Fe)Sb (sample B), when a magnetic field ***H*** was applied in the film plane along the [110] axis (red solid circles) and perpendicular to the plane along the [001] axis (black open circles).

3. **Magnetic field direction ($\theta_H$) dependence of ferromagnetic resonance**

Figures S3(a) and (b) show the FMR spectra for sample A [(Ga,Fe)Sb with BiSb] and sample B [(Ga,Fe)Sb without BiSb], respectively, measured at various magnetic field directions ($\theta_H$) at 300 K. In both samples, the resonance field $\mu_0 H_R$ of the FMR spectra changed from 290 mT to 320 mT when $\theta_H$ is changed from ±90° (***H*** // [1$\bar{1}$0] and [$\bar{1}$10]) to 0° (***H*** // [001]), respectively. As shown in Fig. S3, the FMR spectra and peak positions are almost the same before and after depositing BiSb, thus the magnetic anisotropy remains the same in sample A and sample B. This indicates that the ferromagnetic properties of (Ga,Fe)Sb are not affected by anti-sputtering or BiSb overgrowth.

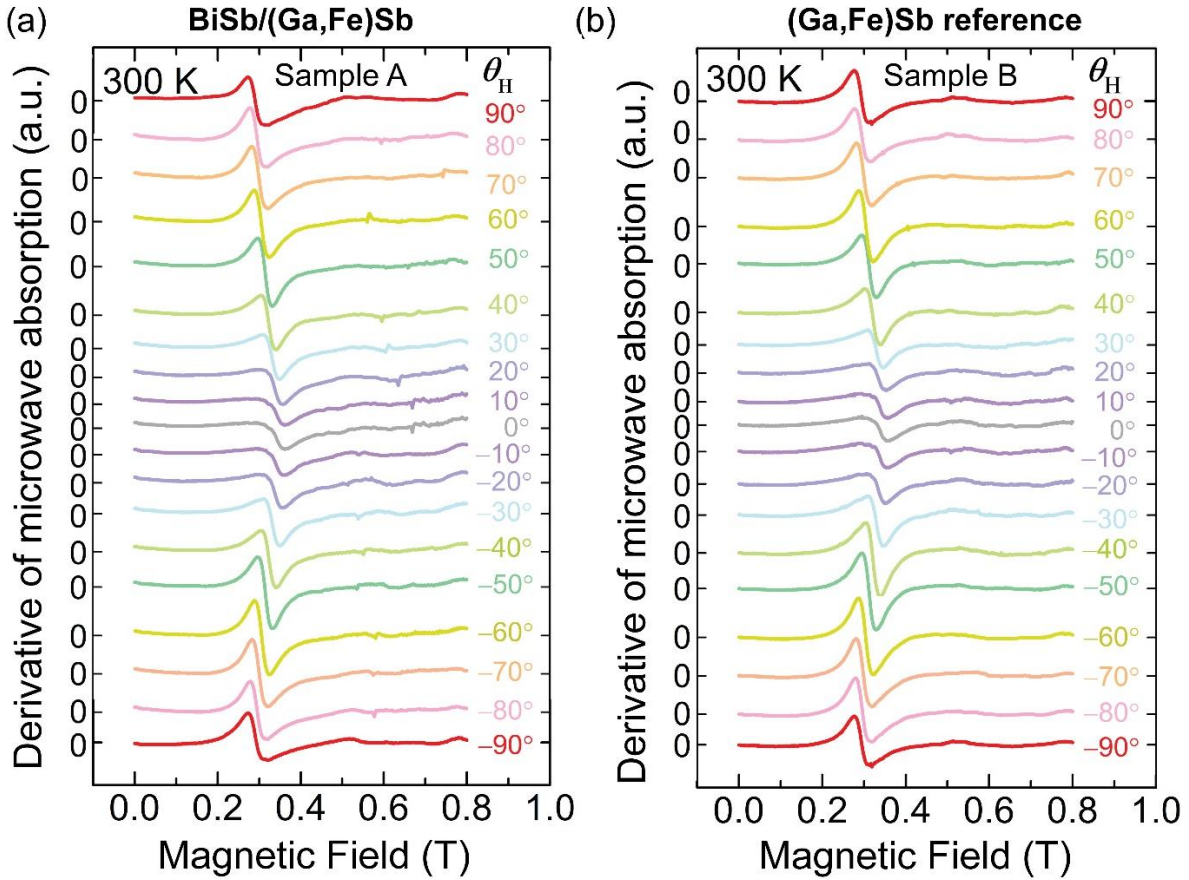

Figure S3. Ferromagnetic resonance (FMR) spectra at 300 K of (a) sample A [(Ga,Fe)Sb with BiSb], and (b) sample B [(Ga,Fe)Sb without BiSb], at various magnetic field directions ($\theta_H$). The used microwave power is 200 mW. $\theta_H$ is defined in Fig. S4 or Fig. 1 (e) in the main text.

## 4. Spin pumping measurement

In spin pumping experiments, we used a Bruker electron paramagnetic resonance (EPR) spectrometer whose cavity resonates in the transverse electric (TE$_{011}$) mode with a microwave frequency $f$ of 9.1 GHz (X-band). We cut the samples into a 3 × 1 mm$^2$ piece with edges along the [110] (3 mm) and [1$\bar{1}$0] (1 mm) axes. For electrical measurements, we have connected the two gold wires to the indium contacts at both edges of the sample with a distance of $l = 2$ mm apart as shown in Fig. S4 (same as Fig. 1(e) in the main text). The sample was placed on the center of a quartz rod and inserted in the center of the microwave cavity. For the measurements, a static magnetic field $\mu_0 H$ is applied along the [1$\bar{1}$0] **axis** in the film plane (except for the measurements with varying $\theta_H$), which corresponds to the easy magnetization axis of (Ga,Fe)Sb. A microwave magnetic field $h$ was applied along the [110] axis [Fig. S4 and Fig. 1(e) in main text]. Also, an ac modulation field $H_{ac}$ (1 mT, 100 kHz) parallel to $H$ is superimposed to obtain the FMR spectrum in its derivative form. The voltage between the indium contacts is detected by a nano-voltmeter. We have measured the FMR signal and the voltage signal $V$ between the indium contacts simultaneously by sweeping the magnitude of $H$. When the FMR condition is satisfied, a pure spin current with the spin polarization parallel to the magnetization precession axis in the (Ga,Fe)Sb layer is injected into the BiSb layer by spin pumping. This injected spin current is converted to a charge current by ISHE, which generates a voltage signal between the edges in the BiSb layer. As discussed in the main text, this voltage signal sometimes contains galvanomagnetic effects, which is originated due to the shift in the sample alignment from the center position inside the cavity. As a result, the microwave electric field produces galvanomagnetic effects. These galvanomagnetic effects can be separated from the inverse spin Hall effect (ISHE) signal by decomposing the voltage signals into symmetric and asymmetric components using Eq. (1) in the main text. In the case of the ideal microwave cavity, i.e. the 90° phase difference between the radio frequency (rf) electric field and the rf magnetic field, the symmetric component consists of the planar Hall effect (PHE) and ISHE, while the asymmetric component composed of only the anomalous Hall effect (AHE). However, the situation becomes complicated if the phase difference is shifted from 90°. Thus, to avoid such complications, we assumed an ideal microwave cavity in this work.

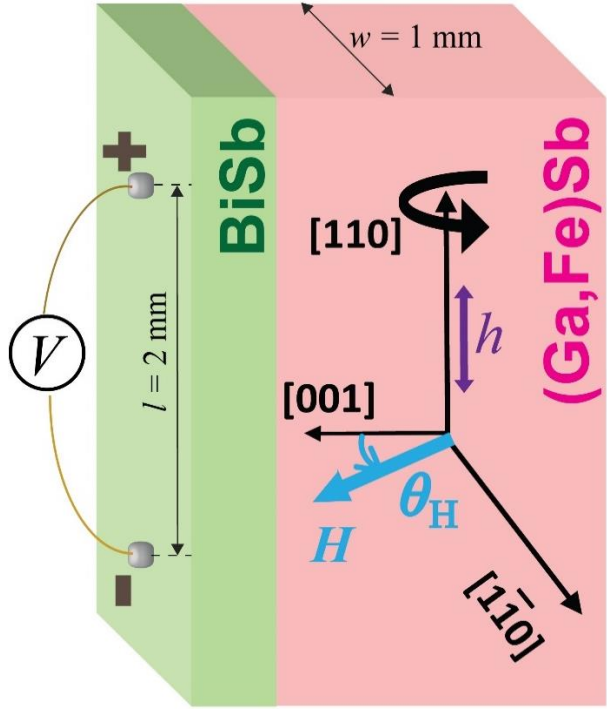

Figure S4. Sample alignment and coordinate system used in the spin pumping measurement. A microwave magnetic field $h$ was applied along the [110] axis of the sample. $\theta_H$ is the angle of the magnetic field $H$ with respect to the [001] axis, respectively, where $H$ is in the (110) plane. $l$ is the distance between the two gold wires connected to the indium contacts.

5. **Estimation of spin current density and spin mixing conductance:**

We estimated the spin current density in sample A by using the following equation:

$$j_S^{\text{BiSb/GaFeSb}} = \frac{g_r^{\uparrow\downarrow} \gamma^2 h^2 \hbar \left( \mu_0 M_S \gamma + \sqrt{(\mu_0 M_S \gamma)^2 + 4\omega^2} \right)}{8\pi\alpha^2 \left[ (\mu_0 M_S \gamma)^2 + 4\omega^2 \right]}, \tag{S1}$$

Here, $\gamma = g\mu_B/\hbar$ is the gyromagnetic ratio, where $g$, $\mu_B$, and $\hbar$ are the $g$ factor, Bohr magneton, and Dirac's constant, respectively, and $\alpha = \sqrt{3}\gamma\Delta H_{BiSb/GaFeSb}/2\omega$ is the damping constant, and $\Delta H_{BiSb/GaFeSb}$ (~ 35 mT) is the FMR spectral linewidth of sample A. $h$, $\mu_0 M_S$, and $g_r^{\uparrow\downarrow}$ are the microwave magnetic field, saturation magnetization, and real part of the spin mixing conductance, respectively. The real part of the spin mixing conductance is calculated by

$$g_r^{\uparrow\downarrow} = \frac{2\sqrt{3}\pi\mu_0 M_S \gamma d_{GaFeSb}}{g\mu_B \omega}(\Delta H_{BiSb/GaFeSb} - \Delta H_{GaFeSb}), \quad (S2)$$

where $\Delta H_{GaFeSb}$ (~ 32 mT) and $d_{GaFeSb}$ are the FMR spectral linewidth of the sample B and the thickness of the (Ga,Fe)Sb film, respectively. Using Eq. (S1), Eq. (S2), and the parameters shown in Table SI, we estimated the value of $g_r^{\uparrow\downarrow}$ and spin current density $j_S^{BiSb/GaFeSb}$ at the BiSb/(Ga,Fe)Sb interface at 300 K. The estimated values of $g_r^{\uparrow\downarrow}$ and $j_S^{BiSb/GaFeSb}$ are $1.24 \times 10^{19}$ m$^{-2}$ and $5.7 \times 10^{-12}$ J m$^{-2}$, respectively.

TABLE SII. The parameters obtained by fitting the model to the experimental FMR spectra. $\Delta H_{BiSb/GaFeSb}$ and $\Delta H_{GaFeSb}$ are the spectral linewidths for sample A and sample B, respectively; $h$, $\mu_0 H_R$, $\omega$, and $g$ are the microwave magnetic field, resonance field, saturation m angular microwave frequency, and $g$ factor, respectively. $\mu_0 M_S$ is the saturation magnetization measured by SQUID.

| $\Delta H_{BiSb/GaFeSb}$ (mT) | $\Delta H_{GaFeSb}$ (mT) | $h$ (T) | $\mu_0 H_R$ (mT) | $\mu_0 M_S$ (mT) | $\gamma$ (T$^{-1}$ s$^{-1}$) | $\omega$ (s$^{-1}$) | $g$ factor |
|---|---|---|---|---|---|---|---|
| ~35 | ~32 | $5.5 \times 10^{-5}$ | 290 | 56 | $1.86 \times 10^{11}$ | $5.74 \times 10^{10}$ | $2.05 \pm 0.01$ |